\documentclass[preprint,12pt]{elsarticle}

\usepackage{dcolumn}
\usepackage{bm}
\usepackage{graphicx}
\usepackage{etoolbox}
\usepackage[dvipsnames]{xcolor}
\usepackage{color}
\usepackage{hyperref}
\usepackage{latexsym}
\usepackage{amsthm}
\usepackage{amsmath}
\usepackage{amssymb}
\usepackage{verbatim}
\usepackage{setspace}
\usepackage{wasysym}
\usepackage{tikz}
\usepackage{tabu}
\usetikzlibrary{shapes}
\usepackage[english]{babel}
\usepackage[utf8]{inputenc}

\DeclareRobustCommand{\uvec}[1]{{%
  \ifcsname uvec#1\endcsname
     \csname uvec#1\endcsname
   \else
    \bm{\hat{\mathbf{#1}}}%
   \fi
}}

\newrobustcmd*{\mysquare}[1]{\tikz{\filldraw[draw=#1,fill=#1] (0,0) rectangle (0.15cm,0.15cm);}}

\newrobustcmd*{\mycircle}[1]{\tikz{\filldraw[draw=#1,fill=#1] (0,0) circle [radius=0.075cm];}}

\newrobustcmd*{\mydiamond}[1]{\tikz{\filldraw[draw=#1,fill=#1] (0.0cm,0.1cm) --
(0.075cm,0.2cm) -- (0.15cm,0.1cm) --  (0.075cm,0.0cm) -- (0.0cm,0.1cm) ;}}

\newrobustcmd*{\myhexagon}[1]{\tikz{\filldraw[draw=#1,fill=#1] (0.0cm,0.05cm) --
(0.0cm,0.1cm) -- (0.075cm,0.15cm) --  (0.15cm,0.1cm) -- (0.15cm,0.05cm)-- (0.075cm,0.0cm) -- (0.0cm,0.05cm) ;}}

\newrobustcmd*{\mytriangle}[1]{\tikz{\filldraw[draw=#1,fill=#1] (0,0) --
(0.2cm,0) -- (0.1cm,0.2cm);}}

\newrobustcmd*{\mytriangledown}[1]{\tikz{\filldraw[draw=#1,fill=#1] (0.1cm,0.0cm) --
(0.0cm,0.2cm) -- (0.2cm,0.2cm);}}

\newrobustcmd*{\myplus}[1]{\tikz{\filldraw[draw=#1,fill=#1] (0.0cm,0.08cm) --
(0.0cm,0.12cm) -- (0.08cm,0.12cm)-- (0.08cm,0.2cm)-- (0.12cm,0.2cm)-- (0.12cm,0.12cm)-- (0.2cm,0.12cm)-- (0.2cm,0.08cm)-- (0.12cm,0.08cm)-- (0.12cm,0.0cm)-- (0.08cm,0.0cm)-- (0.08cm,0.08cm)-- (0.0cm,0.08cm);}}

\begin{document}

\begin{frontmatter}

\title{Contact processes on fragmented domains}

\author[inst1]{I. Ibagon}

\ead{ingridibagon@gmail.com}
\address[inst1]{Departamento de F\'isica, ICEx, Universidade Federal de Minas Gerais, C. P. 702, 30123-970, Belo Horizonte, Minas Gerais, Brazil
  }

\author[inst1]{A. P. Furlan}
\ead{apfurlan@fisica.ufmg.br}

\author[inst1,inst2]{Ronald Dickman}
\ead{dickman@fisica.ufmg.br}
\address[inst2]{National Institute 
of Science and Technology for Complex Systems,Brazil}

\date{\today}

\hypersetup{citecolor=blue,
  urlcolor=magenta,
  pdfcreator={pdflatex},
}

\begin{abstract} 
  Motivated by recent findings of enhanced species survival when fragmented habitats are reconnected through narrow strips of land [S. Pimm, and C. N. Jenkins, Am. Sci. {\bf 107}(3), 162 (2019).], we study the effect of a corridor connecting two square regions on the survival time of three models exhibiting extinction/survival phase transitions: the basic contact process (CP),  the diffusive contact process, and the two-species symbiotic contact process (2SCP). We find that connecting the regions generally increases the survival time for $\lambda\ge \lambda_c$, where $\lambda$ is the reproduction rate and $\lambda_c$ its critical value. The enhancement of the survival time increases with $\lambda$, and  is largest in the 2SCP.
 
\end{abstract}

\end{frontmatter}
\section{Introduction}
Human activities such as construction and agriculture are responsible for habitat loss and fragmentation \cite{MEA}. An alarming consequence of fragmentation is that species richness decreases, with smaller fragments losing species more rapidly \cite{Pimm, Haddad15}.
Since habitat fragments---even when protected--- are often too small to maintain viable populations of species, corridors connecting the fragments are used as a conservation strategy \cite{CE, Pimm}. Most studies of the effect of corridors on populations are empirical. Beier and Noss  \cite{BandN} reviewed 32 studies published from 1988 to 1997 and concluded that  they provide evidence on the benefit of corridors to populations in real fragmented landscapes, increasing immigration and colonization rates or facilitating movement of individuals between fragments. Similarly, a meta-analysis of 78 experiments published between 1988 and 2008 found a 50\% increase in movement between habitat patches connected with corridors in comparison with habitat patches without corridors \cite{Norton2010}. Haddad et al. \cite{Haddad2014} reviewed studies on the negative effects of corridors, such as increased dispersal of antagonistic species to the conservation target, increased dispersal of exotic species, spread of disturbances like fire, and  edge effects. However, no consistent evidence of negative effect of corridors was found.

Theoretical work in this area includes a two-population model\cite{Hu2003} and metapopulation models \cite{Hanski2000,Ov2012}. In Ref. \cite{Hu2003}, two logistically growing populations within two patches are connected by migration, which depends on the presence or absence of the corridor\cite{Hu2003}. In spatial metapopulation models, the dynamics of the fraction of occupied patches, representing local populations, is governed by colonization and extinction rates that depend on the patch area and the distance between patches \cite{Hanski2000, Ov2012}. Given the growing use of corridors in conservation ecology we believe the area might benefit from analysis of simple models with explicit spatial structure, which may help to understand which characteristics of the corridor and/or the population are important for this strategy to succeed. In this work we study the effect of linking fragmented regions on species survival times in the contact process and related models.

The contact process (CP) \cite{Harris} is a simple stochastic system that can be interpreted as a birth-and-death process with spatial structure. Since the CP undergoes a phase transition between extinction and survival as the reproduction rate $\lambda$ is varied, it is a reasonable first choice to model the effect of connecting habitable fragments. In addition, the CP is readily adapted to model other factors affecting population dynamics, such as disorder\cite{ARD96,RAD98,Vo2009} or symbiotic interactions \cite{MMO2012,MMO2014,SCP2018}.

Here, we study the effect of a corridor linking two square regions on the survival time of the basic CP, the diffusive CP, and the two-species symbiotic CP (2SCP)\cite{MMO2012}. The rest of this paper is organized as follows. In Sec. \ref{sec:models} we introduce the models, in Sec. \ref{sec:CP-CG} we present a semianalytic solution of the basic CP on a {\it complete graph}, and in Sec. \ref{simulations} present our simulation results. We conclude and discuss our findings in Sec. \ref{cd}.

\section{Models}\label{sec:models}

The basic CP \cite{Harris} is a Markov process  defined on a lattice. Each lattice site $i$ is either occupied(active) [$\sigma_i=1$] or vacant(inactive) [$\sigma_i=0$]. Transitions from $\sigma_i=0$ to 1 occur at a rate $r\lambda$, where $r$ is the fraction of nearest neighbors (NNs) of site $i$ that are occupied and $\lambda$ is the reproduction rate. The transition from $\sigma_i=1$ to 0 is independent of the neighboring sites and occurs at a rate of unity (Thus, our time unit corresponds to the mean life of an organism). The state $\sigma_i=0$ for all $i$ is absorbing. The model exhibits a phase transition from the active to the absorbing state at a certain critical value $\lambda_c$; for the two-dimensional square lattice studied here $\lambda_c=1.64877(3)$ \cite{RD99}. In finite systems, the only stationary state is the absorbing state, and one therefore studies the {\it quasistationary state} which describes the statistical properties of surviving realizations at long times. After an initial stage whose duration depends on the system size $L$, and on $\Delta=\lambda-\lambda_c$, averages over surviving realizations attain steady values. Considering the evolution starting from a fully occupied lattice, the survival time  $\tau(\Delta,L)$ is defined such that the survival probability $P(t)\propto \exp{[-t/\tau(\Delta, L)]}$ for long times $t$. At the critical point ($\Delta=0$), $\tau(0,R)\sim L^{\nu_{\parallel}/\nu_{\perp}}$, where $\nu_{\parallel}$ and $\nu_{\perp}$ are critical exponents. In the active phase ($\Delta>0$), the scaling form for $\tau(L, \Delta)$ in $d$ dimensions is given by\cite{RAD98},

\begin{equation}\label{Eq:tau}
   \tau(\Delta,L)\sim  L^{\nu_{||}/\nu_{\perp}}\exp [c(L\Delta^{\nu_{\perp}})^{d}], 
\end{equation}
 where $c>0$ is a constant. Thus the survival time grows exponentially with the number of sites in the supercritical regime. In the subcritical regime ($\Delta<0$), the survival time is expected to follow \cite{MMO2006},
\begin{equation}
   \tau(\Delta, L)\sim\left| \Delta \right|^{\nu_{\parallel}}\mathcal{F}\left(\left|\Delta\right|L^{1/\nu_\perp}\right), 
\end{equation}
where the scaling function $\mathcal{F} (x) \propto  x^{\nu_\parallel}$ for small $x$. In this regime, the survival time approaches a constant value as $L\to \infty$ .

In addition to the basic CP, we study two variants, each of which includes features relevant to real populations and/or habitats: (1) The diffusive contact process in which, in addition to creation and death, each individual attempts to hop to one of its NN sites at rate $D$ \cite{Liggett,Je93}; and (2) the two-species symbiotic CP (2SCP)\cite{MMO2012}. 

In the 2SCP, two species A and B, can occupy the sites of the lattice. The allowed states for a site $i$, ($\sigma_i$, $\eta_i$), are (0,0), (0,1), (1,0) and (1,1), where $\sigma_i$ and $\eta_i$ indicate the occupation of site $i$ by species A and B, respectively. Transitions (0,1) to (0,0) and (1,0) to (0,0) occur at a rate of unity, while (1,1) to (1,0) or to (0,1) occur at a rate $\mu$. Transitions (0,0) to (1,0) and (0,1) to (1,1) occur at a rate $r_A\lambda$, where $r_A$ is the fraction of NNs containing a particle of species $A$. Similarly, transitions (0,0) to (0,1) and (1,0) to (1,1) occur at a rate $r_B\lambda$ with $r_B$ the fraction of NNs containing a particle of species $B$. For $\mu=1$, this set of transition rates describes a pair of independent contact processes cohabiting the lattice, but for $\mu<1$ the annihilation rate is reduced at sites with both species, representing symbiosis between the species. Details of the simulation algorithm can be found in Ref. \cite{MMO2012}.

The nondiffusive models better describe sessile organisms such as plants. We interpret one lattice site as corresponding to the area needed to support a single organism. For example, if we use the largest value for the density of {\it Eucalyptus salmonophloia} reported in Ref. \cite{No95}, i.e., 170.6 trees per ha, one site of our simulation corresponds to an area of approx. 60~m$^2$, so that a square of 40$\times$40 represents an area of about 0.1 km$^2$.  

\section{Contact process on a complete graph}\label{sec:CP-CG}

Since the CP is not exactly soluble on finite-dimensional lattices, a reliable approach to determine the effect of domain connection on lifetime is via Monte Carlo simulation.  
Before turning to simulation, we note that a
semianalytic (``numerically exact'') solution is possible for the CP on a 
{\it complete graph}, that is, a graph of $N$ nodes, all of which which are neighbors.  This can be seen as representing a situation in which organisms are highly mobile, visiting all parts of
their habitat on a time scale much shorter than their mean lifetime.
In this case the state of the system is fully specified by $n$, the number of active sites.
The critical reproduction rate is $\lambda_c = 1$.
The quasistationary probability distribution $Q(n)$ is readily determined via an iterative
procedure \cite{QSS}.  Then the QS lifetime is $\tau = 1/Q(1)$.  (In the QS distribution,
$Q(0)$ is zero by definition since the process has never visited the state $n=0$.)

To study the effect of connecting separate domains, we consider a pair of complete graphs, of 
size $N_1$ and $N_2$ and a single node A, which represents the corridor.  The latter is connected to all nodes in each of the
complete graphs.  Thus an organism in region 1 can produce an offspring in A, which can then
produce an offspring in region 2.  (Note that there are no direct connections between regions
1 and 2.)  

Let $\tau_1$ and $\tau_2$ be the mean survival times of the CP populations
restricted to their respective regions when there is no corridor between them.  The mean
survival time of the {\it composite} region is max$[\tau_1, \tau_2] \equiv \tau_0$.  
If we connect the two regions via site A, the survival time grows to $\tau_c$.  In the case of complete graphs 
1 and 2 connected by node A, we can again solve for the quasistationary probability
distribution $Q(n_1,n_2,n_A)$, where $n_1$ ranges from zero to $N_1$ and similarly
for $n_2$, and $n_A$ is either zero or one.  The QS survival time of the connected system
is

\begin{equation}
\tau_c = \frac{1}{Q(1,0,0) + Q(0,1,0) + Q(0,0,1)}.
\end{equation}  
\vspace{.5em}

\begin{figure}[h!]
  \centering
  \includegraphics[scale=0.55]{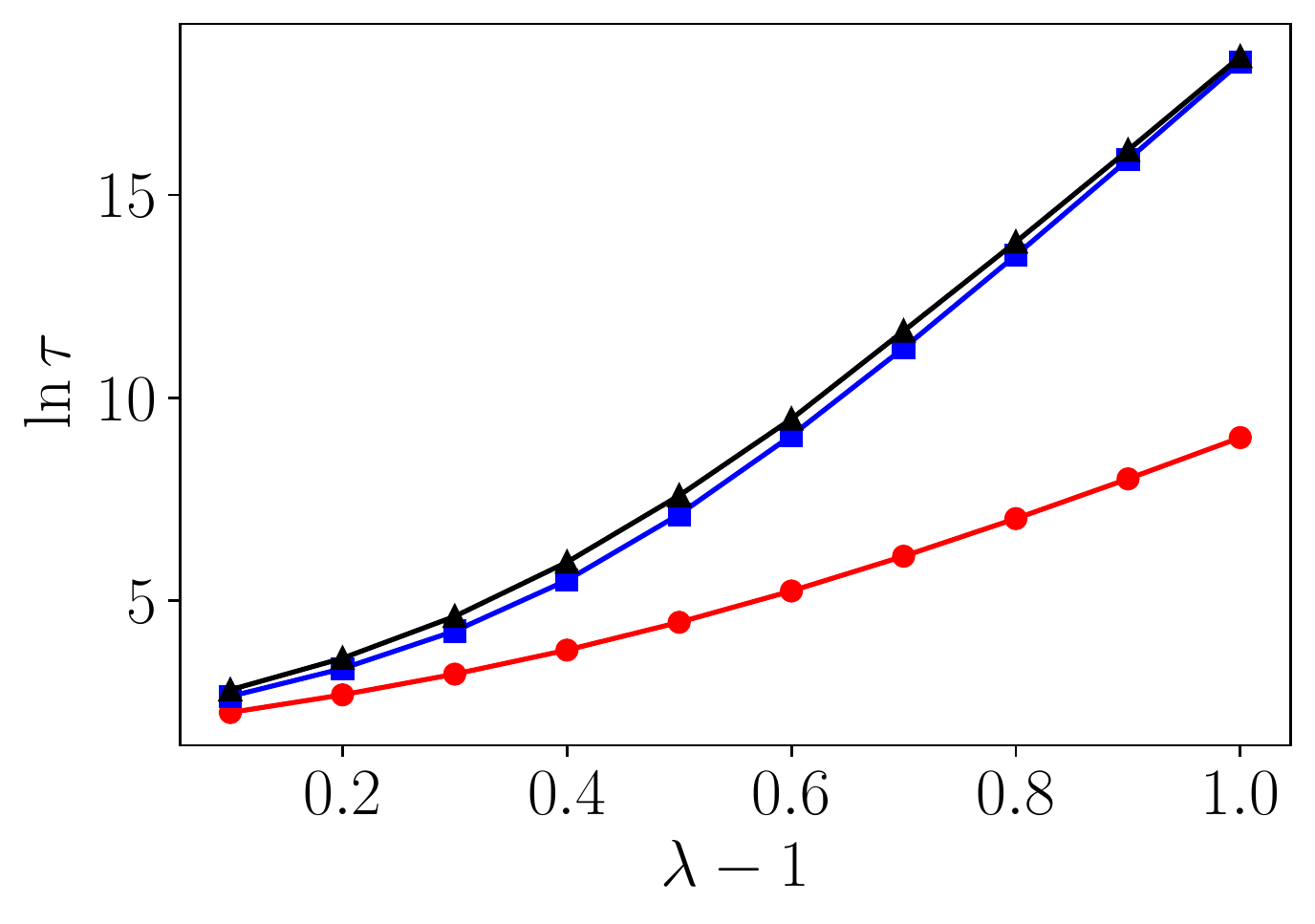}
  \caption{Contact process on a complete graph: Quasistationary survival time on a pair of disconnected complete graphs of 50 nodes (red circles), on a similar pair connected by a single site (blue squares), and on a complete graph of 101 nodes (black triangles).}
  \label{tau1}
\end{figure}

Figure \ref{tau1} shows, for a complete graph of $N=50$ nodes, the mean survival time $\tau_0$ as a function of $\lambda - \lambda_c$ (red curve), and the survival time $\tau_c$ of the composite system (two complete graphs of 50 nodes linked by a single site - the blue curve).  (Recall that the survival time for two {\it disconnected} graphs of the same size is the same as for a single graph.)  Also plotted is the survival time for a {\it single} complete graph of 101 sites (black curve).  We see that the survival time on the connected system approaches that of the single nonfragmented habitat. 

\begin{figure}[h!]
  \centering
  \includegraphics[scale=0.55]{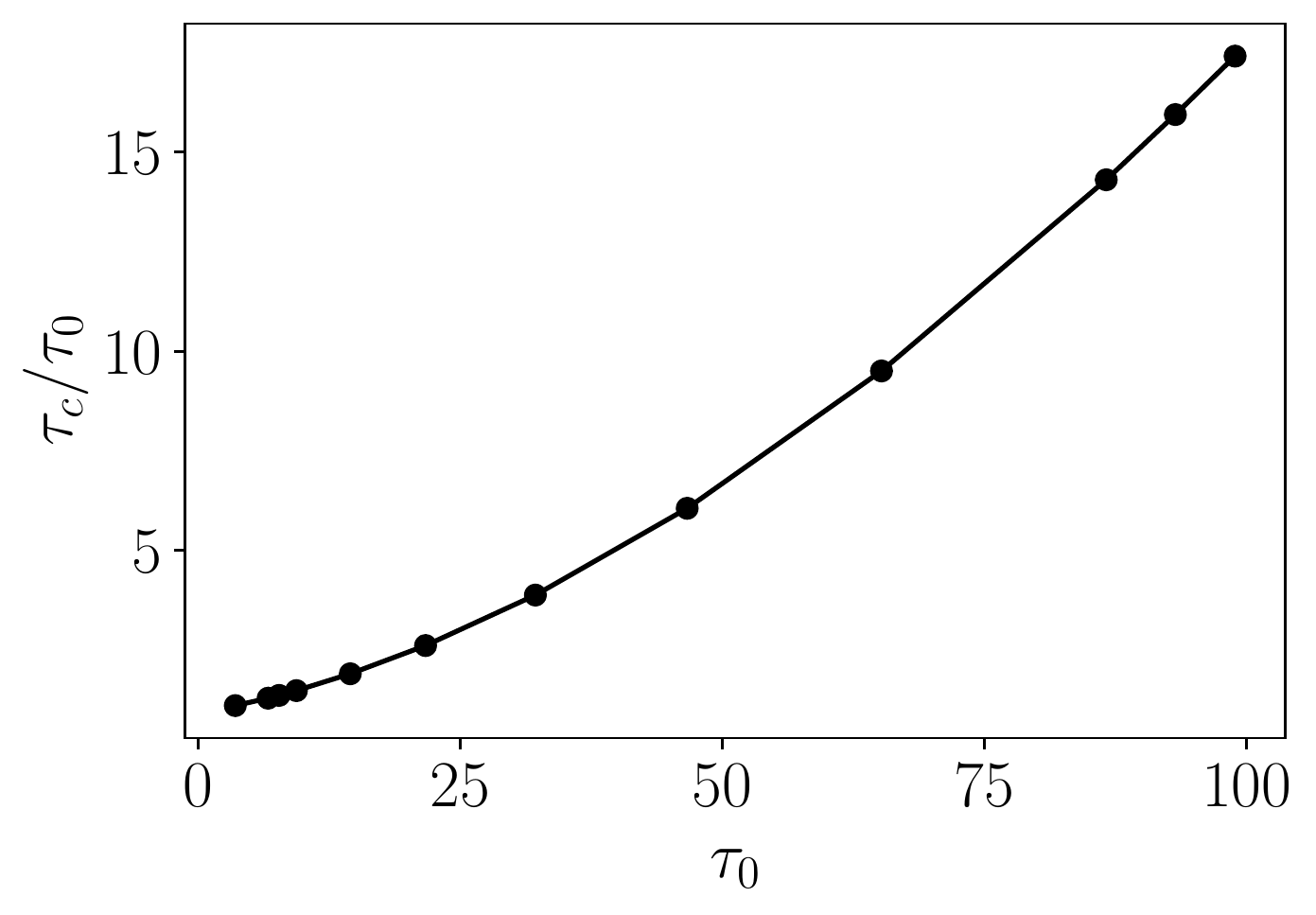}
  \caption{CP on a complete graph: Enhancement ratio $\tau_c/\tau_0$ of the quasistationary lifetime on a pair of graphs of 50 sites connected by a single site to
  that in the absence of the connecting site. }
\label{rattau}
\end{figure}

Figure \ref{rattau} plots the enhancement ratio $\tau_c/\tau_0$ versus $\tau_0$.  For $\tau_0 \leq 100$, 
connecting the two fragments leads to an enhancement in lifetime of up to a factor of nearly 20.

\section{Simulation results}\label{simulations}
\begin{figure}[h!]
    \centering
    \includegraphics[scale=.18]{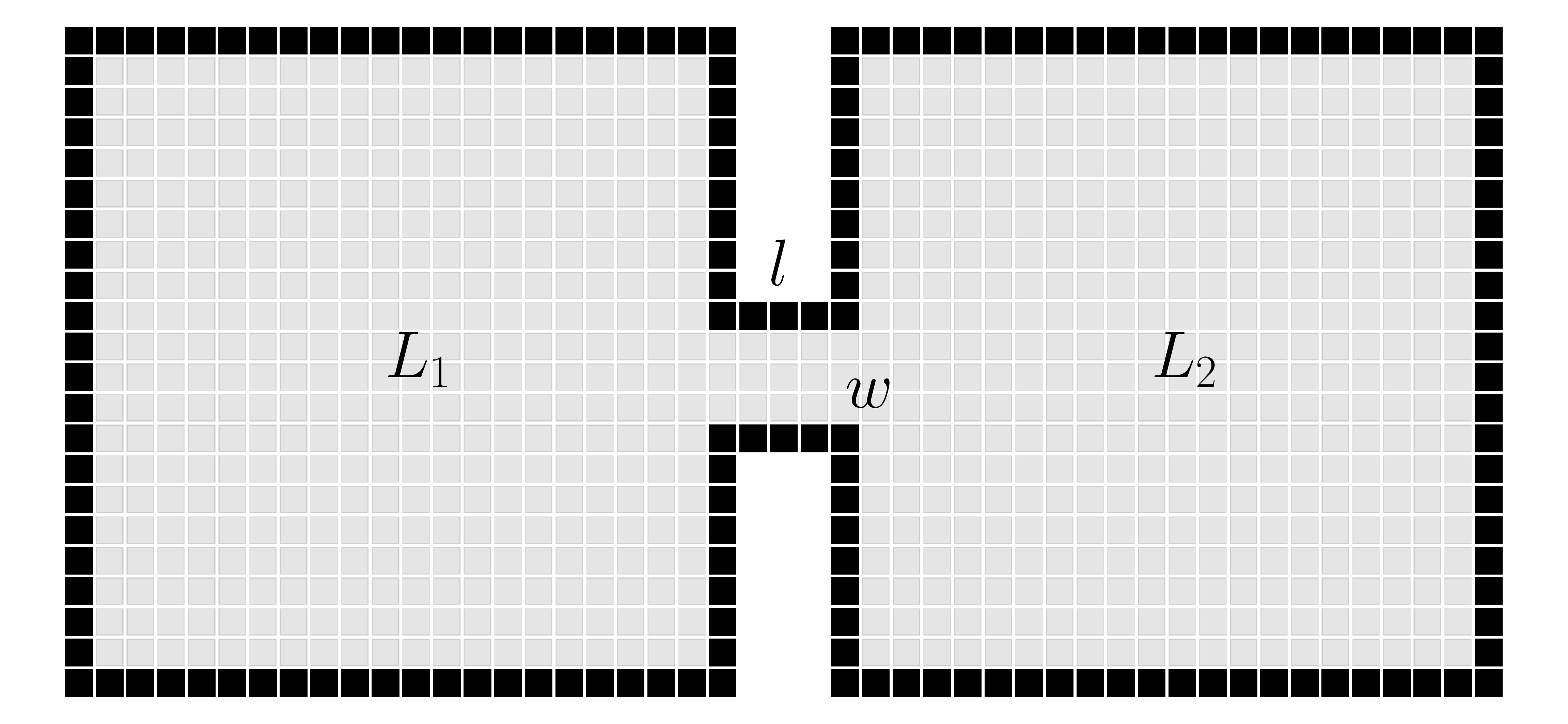}
    \caption{A pair of square regions $L_1=L_2=20$ connected by a link of length $l=5$ and width $w=3$.}
    \label{fig:geometry}
\end{figure}
We perform Monte Carlo simulations of the three models described in Sec. \ref{sec:models} on a pair of square   regions of sides $L_1$ and $L_2$ connected by a corridor of length $l$ and width $w$ (see Fig. \ref{fig:geometry}).
The initial configuration is maximally occupied, i.e., all sites occupied in the CP and the diffusive CP, and all sites doubly occupied in the 2SCP. The sizes $L_1$ and $L_2$  range from 20 to 80; the relative width $\bar w=w/L_1$, and relative length $\bar l=l/L_1$
are varied from $0.05$ to $1.0$. We calculate averages over samples of 50000 independent realizations. The simulations typically required up to 6 hours for each value of $\lambda$ and $L_1=L_2=20$ and up to 350 hours for $L_1=L_2=40$.

\subsection{Contact process}\label{subsec:CP}

\begin{figure}[h!]
    \centering
    \includegraphics[scale=.55]{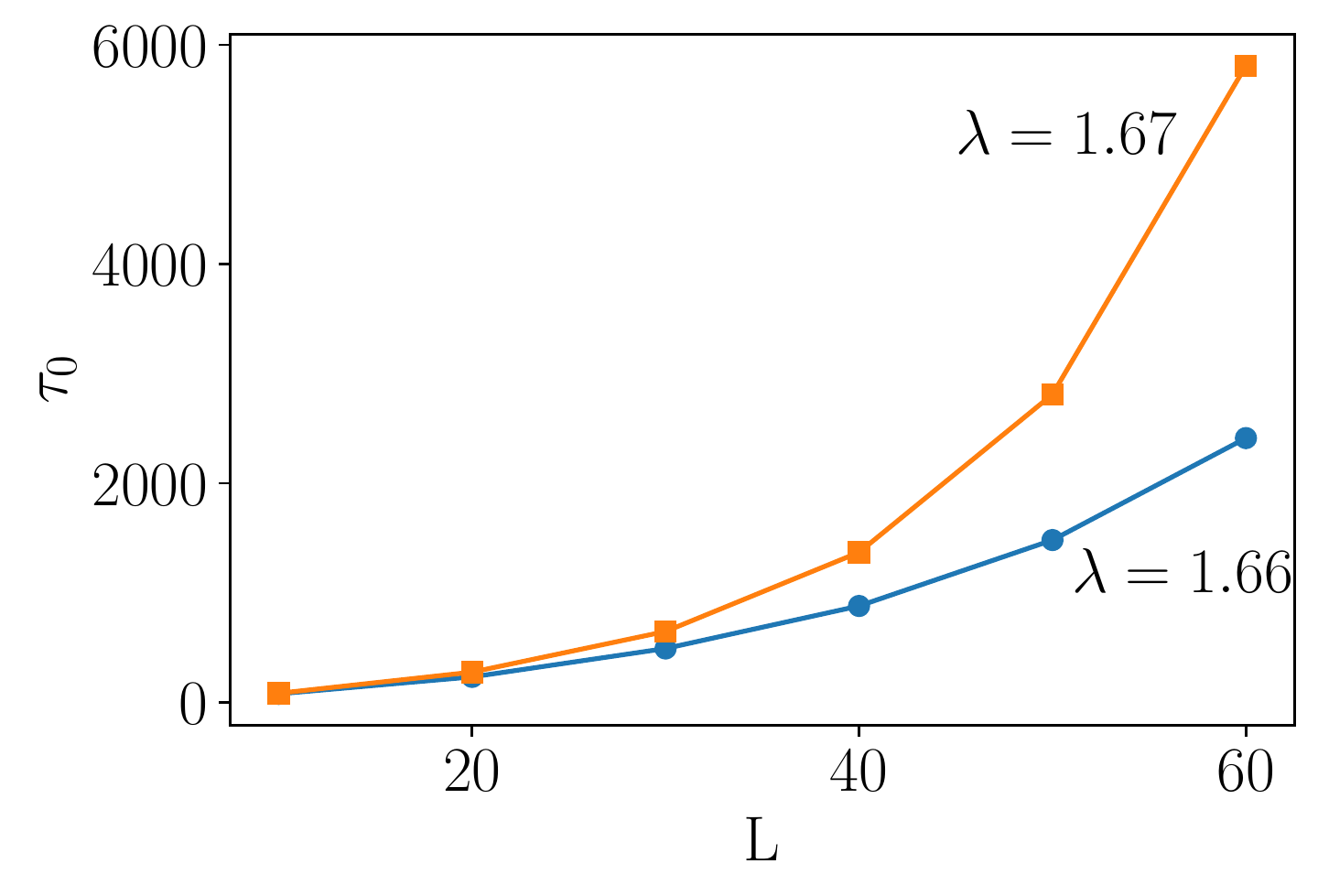}
    
    \caption{ Basic CP: Survival time $\tau_0$ vs system size $L$ for two disconnected square regions $L=L_1=L_2$.}
    \label{fig:tau0}
\end{figure}

\begin{figure}[h!]
    \centering
    \includegraphics[scale=.55]{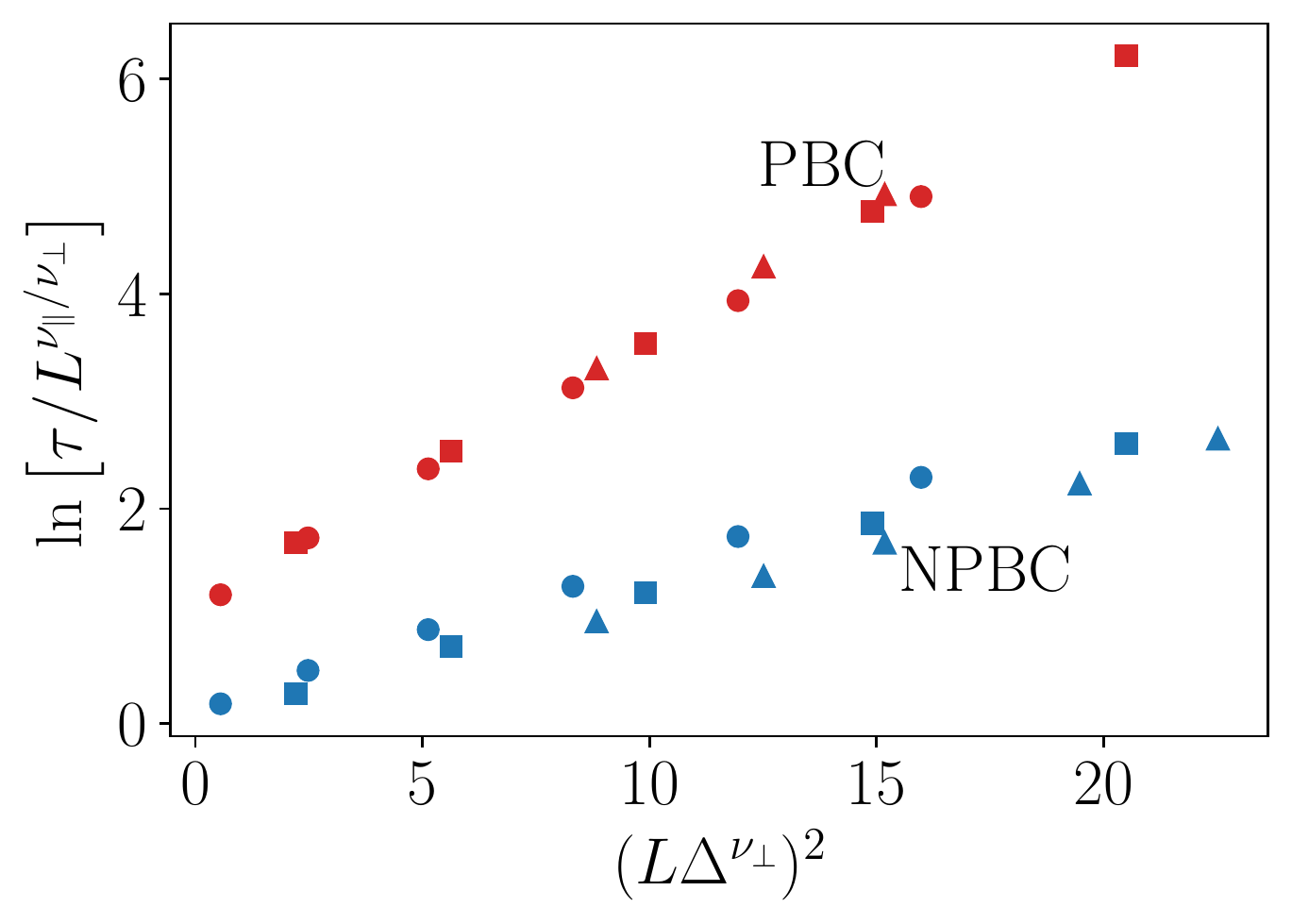}
    
    \caption{ Basic CP: Scaling plot of the survival time $\tau_0$ (see Eg. \ref{Eq:tau})) on a square lattice with periodic boundaries (PBC) and open boundaries (NPBC). System sizes: \mycircle{black} $L=20$, \mysquare{black} $L=40$, $\blacktriangle L=80$. }
    \label{fig:stau0}
\end{figure}

We studied the basic CP for reproduction rate values $\lambda>\lambda_c$ ($\lambda_c=1.64874(4)$). As mentioned in Sec. \ref{sec:models}, following an initial stage during which $P(t)=1$, the survival probability decays exponentially with survival time $\tau$, i.e., $P(t)\propto \exp{[-t/\tau]}$ and the survival time $\tau$ can be obtained from linear fits to $\ln P(t)$. We first determine the survival time $\tau_0$ of two disconnected regions as a function of the system size $L$. Fig. \ref{fig:tau0} shows $\tau_0$ for $L=10-60$ and $\lambda=1.66$ and $1.67$. For fixed reproduction rate, smaller fragments have smaller survival times, as expected.  Since survival times are already large for rather small systems, we focus on system sizes $L=20$ and 40. Figure \ref{fig:stau0} shows scaling plots for $\tau$ on square lattices with periodic boundaries confirming the scaling form in Eq. \ref{Eq:tau}, and $\tau_0$ in the disconnected regions which is equal to the survival time on a square lattice with open boundaries.

\begin{figure}[h!]
    \centering
    \includegraphics[scale=.55]{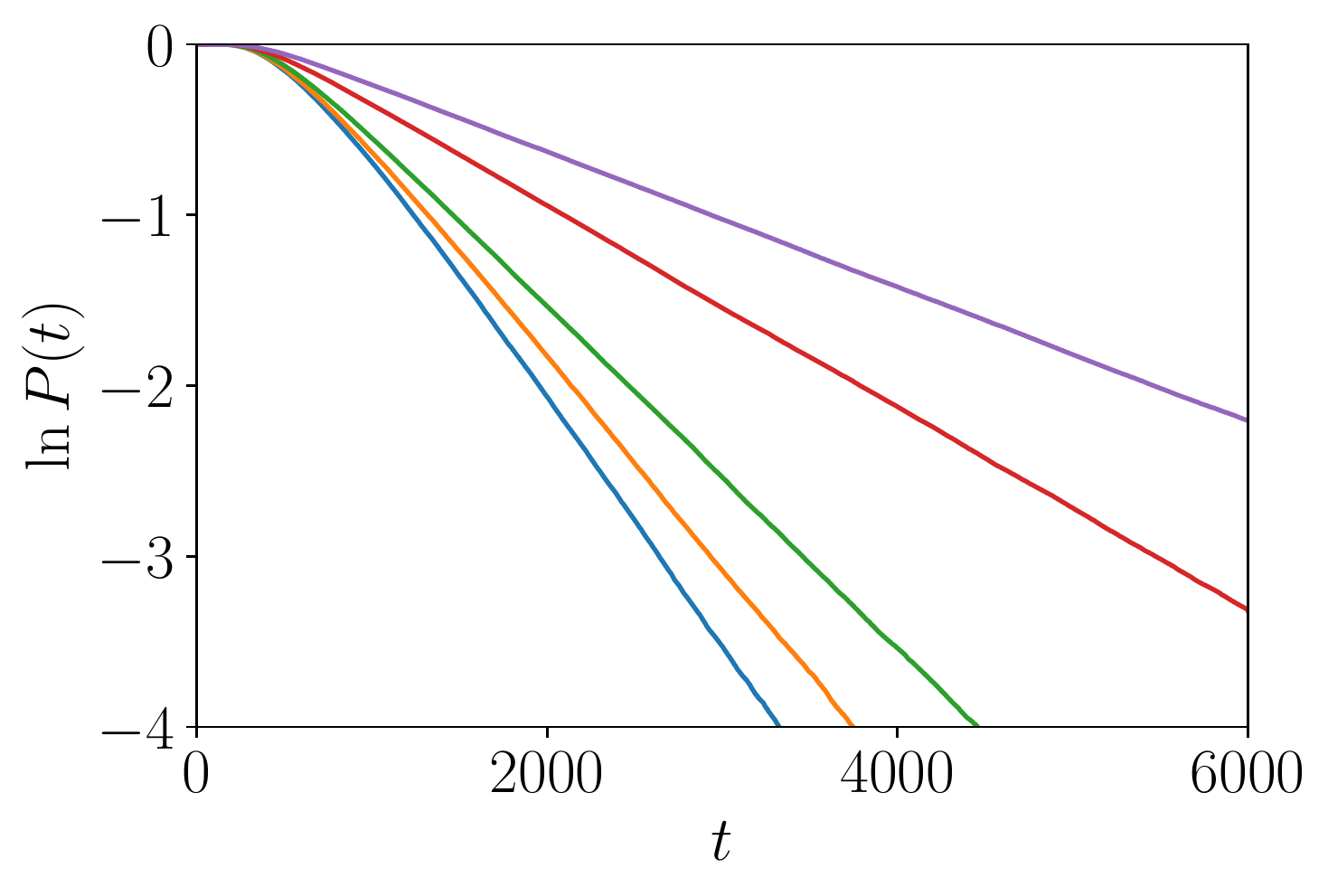}
    \caption{Survival probability $P$ versus time in the basic CP for $\lambda=1.72$ ($\lambda-\lambda_c=0.0712$) on a pair of regions ($L_1=L_2=20$) connected by a corridor of length $\bar l=l/L_1=0.05$ and width (lower to upper) $\bar w=w/L_1=0,0.05,0.15,0.5$ and $1$.($\bar w=0$ corresponds to a habitat consisting of disconnected regions).}.
    \label{fig:P_CPL1}
\end{figure}

Figure \ref{fig:P_CPL1} 
shows a semilogarithmic plot of the the survival probability P(t) for $\lambda=1.72$ for two regions, $L_1=L_2=20$, connected by a corridor of length $\bar l=l/L_1=0.05$ and varying width $\bar w=w/L_1$. Note that $\bar w=0$ corresponds to two disconnected regions and $\bar w=1$ to a rectangular region of $20\times41$ lattice sites.  For fixed reproduction rate $\lambda$ the survival time increases with the corridor width.

\begin{figure}[h!]
    \centering
    \includegraphics[scale=.55]{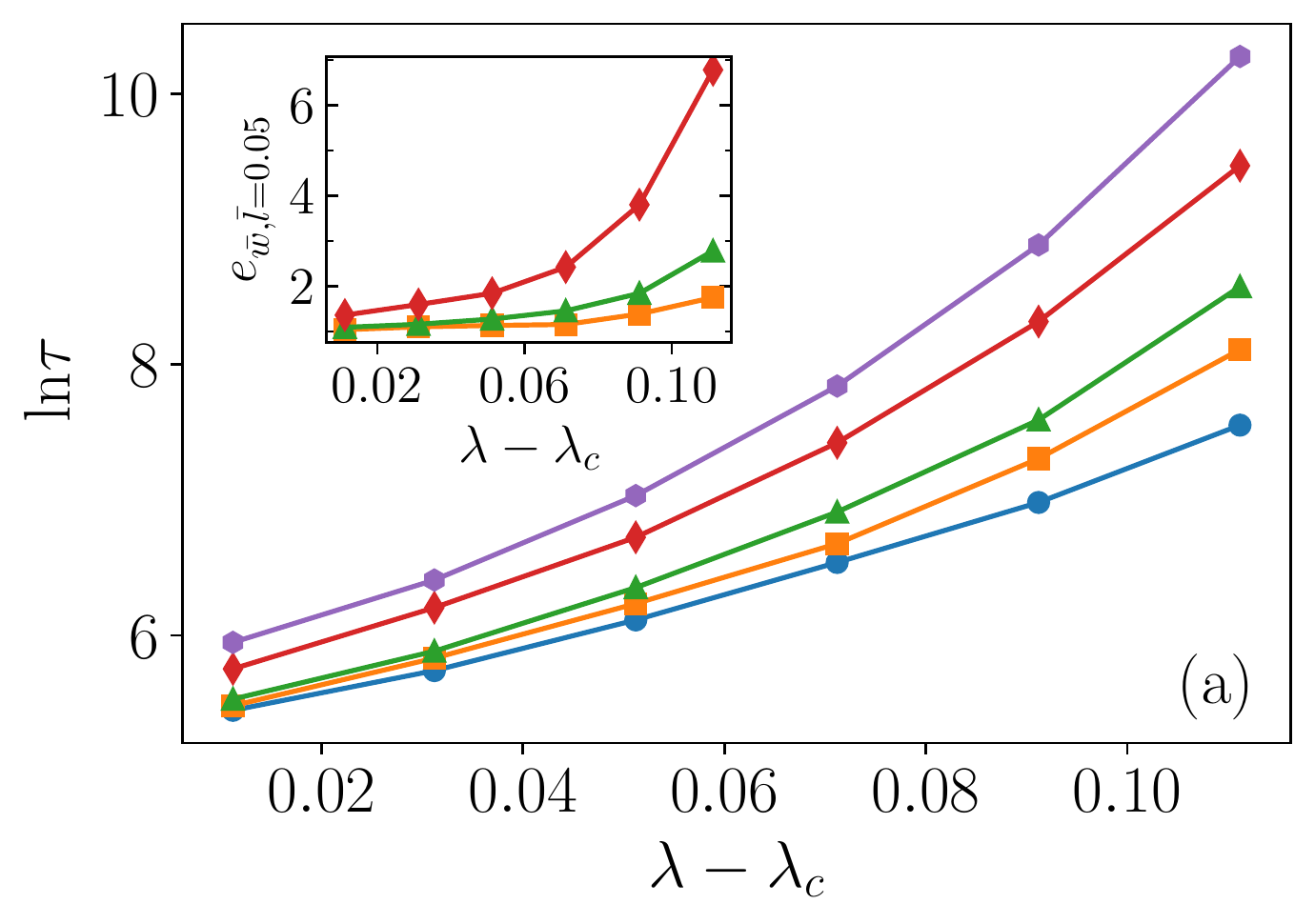}
    \includegraphics[scale=.55]{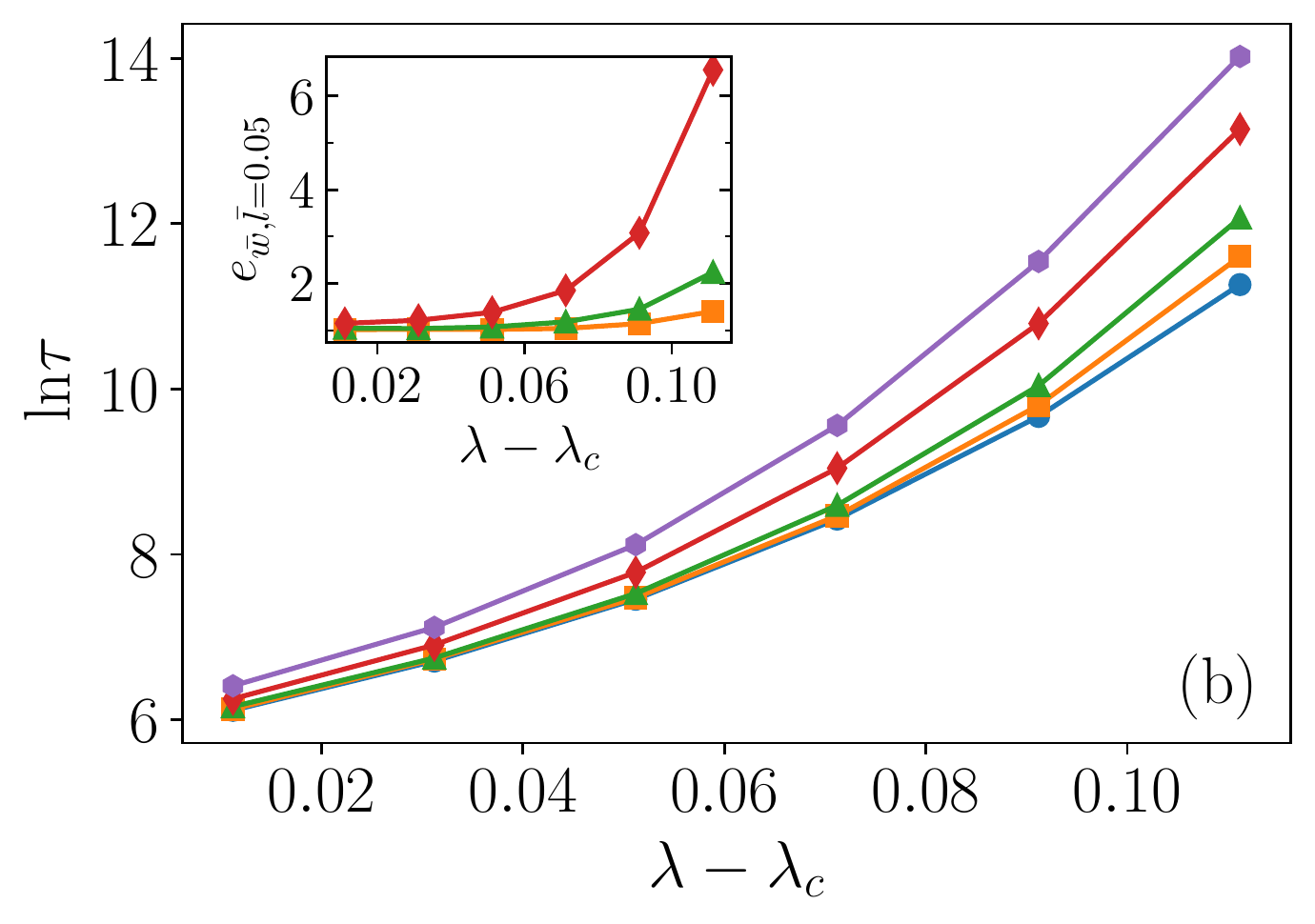}
    \caption{Survival time $\tau$ of the basic contact process on a pair of square regions connected by a corridor of length $\bar l=l/L_1=0.05$ and varying widths $\bar w=w/L_1$ for (a) $L_1=L_2=20$ and (b) $L_1=20,L_2=30$. Symbols: $\bar w=0$ (\mycircle{blue}); $\bar w=0.05$ (\mysquare{Orange}); $\bar w=0.15$  $(\mytriangle{Green})$; $\bar w=0.5$ (\mydiamond{red}); $\bar w=1$ (\myhexagon{Purple}).  Insets show the enhancement ratio $e_{\bar w,\bar l=0.05}=\tau_{\bar w,\bar l=0.05}/\tau_0$.}
    \label{fig:tau_CPL1}
\end{figure}

Figure \ref{fig:tau_CPL1} shows the survival time $\tau$ as a function of $\lambda-\lambda_c$ for two square regions, (a) $L_1=L_2=20$ and  (b) $L_1=20, L_2=30$, connected by a corridor of length $\bar l=0.05$ and varying width $\bar w$. The inset shows the enhancement ratio  $e_{\bar w,\bar l}=\tau_{\bar w,\bar l}/\tau_0$, where $\tau_{\bar w,\bar l}$ is the survival time of a pair of regions connected by a corridor of width $\bar w$ and length $\bar l$, and $\tau_0$ is the survival time of the disconnected regions ($\bar w=0$). Connecting two equal regions can lead to an enhancement in the survival time by up to a factor of 2 for a corridor of width $\bar w=0.05$; the enhancement grows as $\bar w$ is increased. For $L_1=20$ and $L_2=30$, survival times are greater because the systems are larger (see Eq.~(\ref{Eq:tau})), but the enhancement due to the corridor is smaller for $\bar w=0.05$ and $0.15$ (see inset of Fig. \ref{fig:tau_CPL1}(b)). 

\begin{figure}[h!]
    \centering
    \includegraphics[scale=.55]{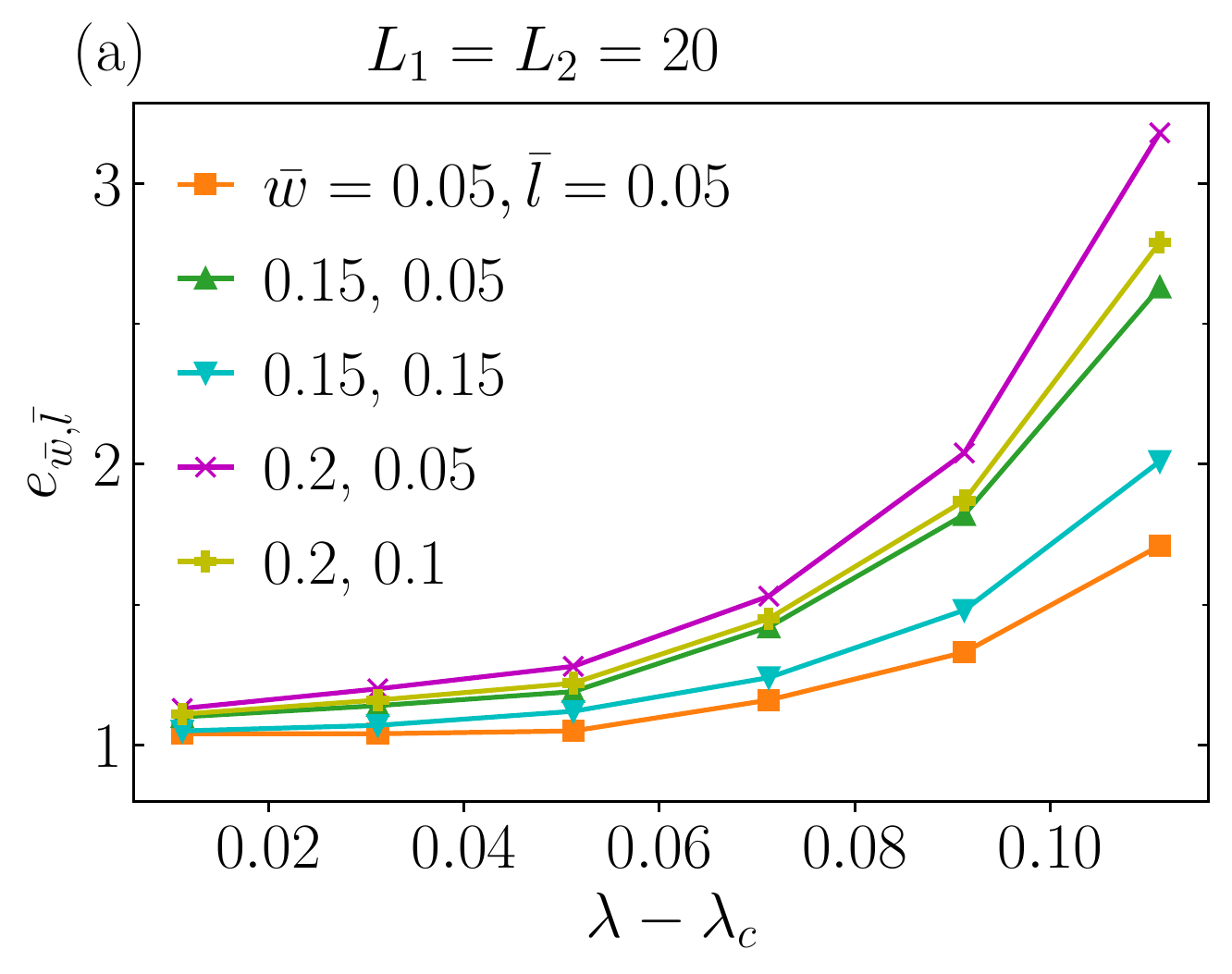}
    \includegraphics[scale=.55]{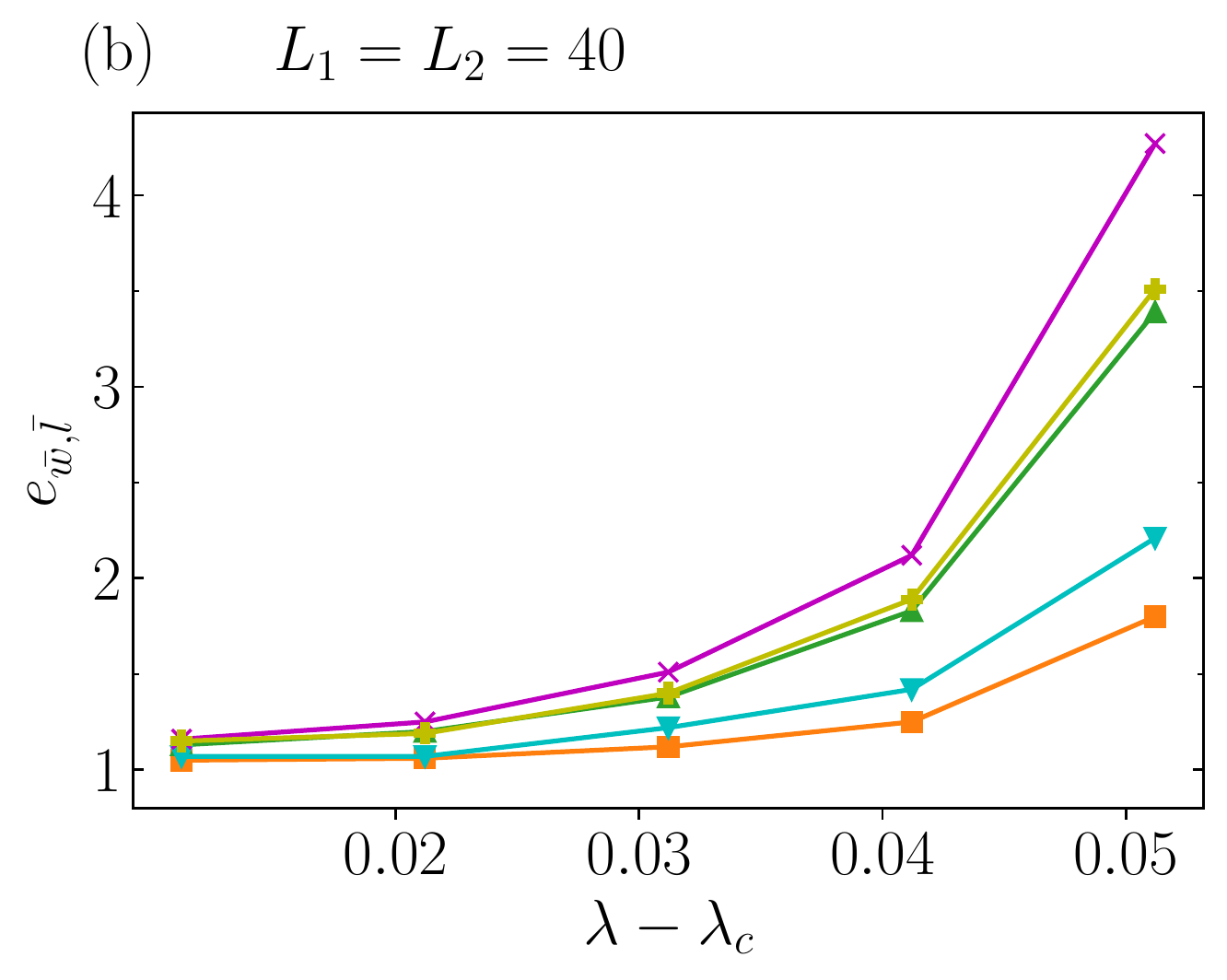}
    \caption{ Basic CP: Enhancement ratio $e_{\bar w,\bar l}=\tau_{\bar w,\bar l}/\tau_0$. 
    (a) $L_1=L_2=20$; (b) $L_1=L_2=40$. Different symbols indicate different corridor geometries, as per the legend in panel (a).}
    \label{fig:rCPdL}
\end{figure}

\begin{figure}
    \centering
    \includegraphics[scale=.45]{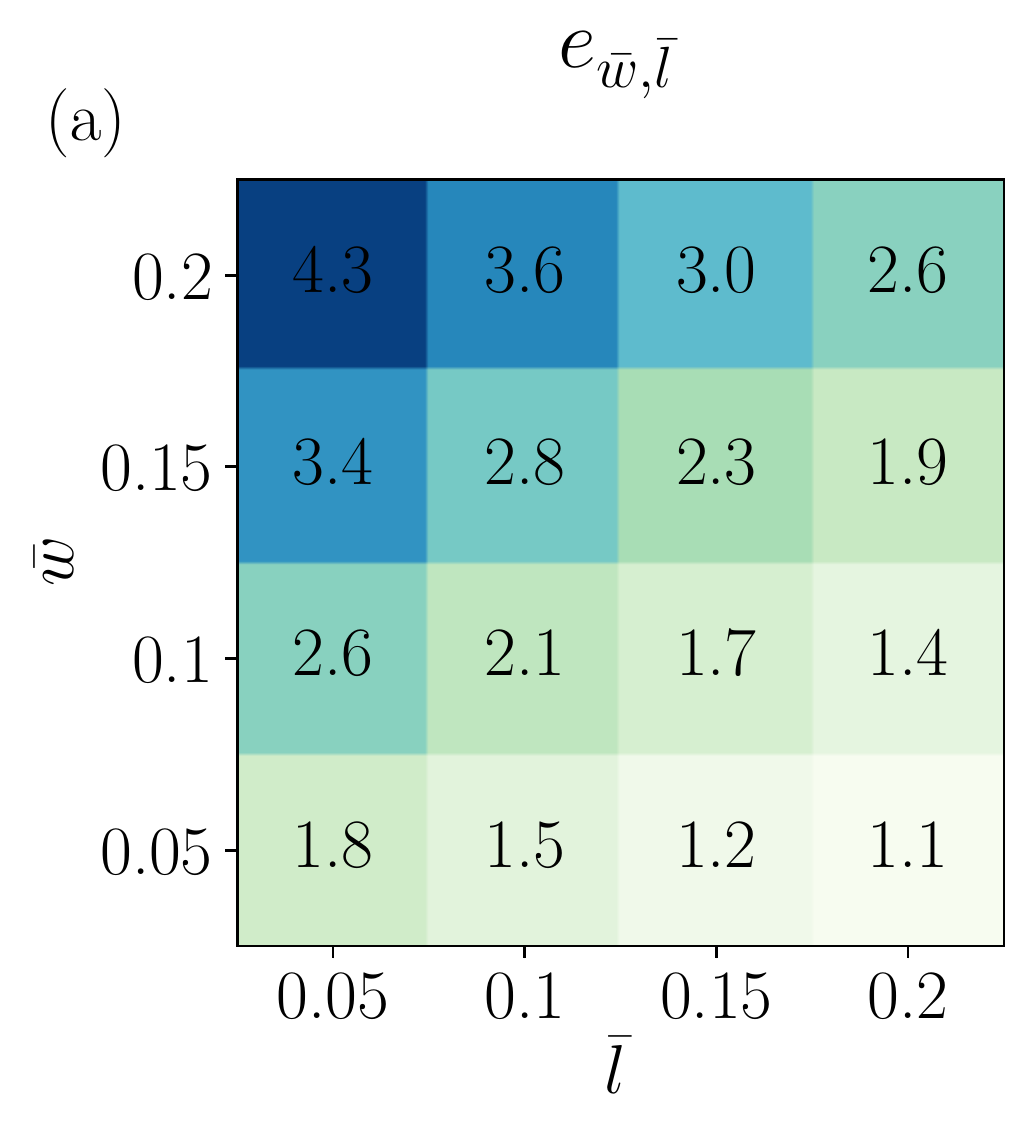}
    \includegraphics[scale=.45]{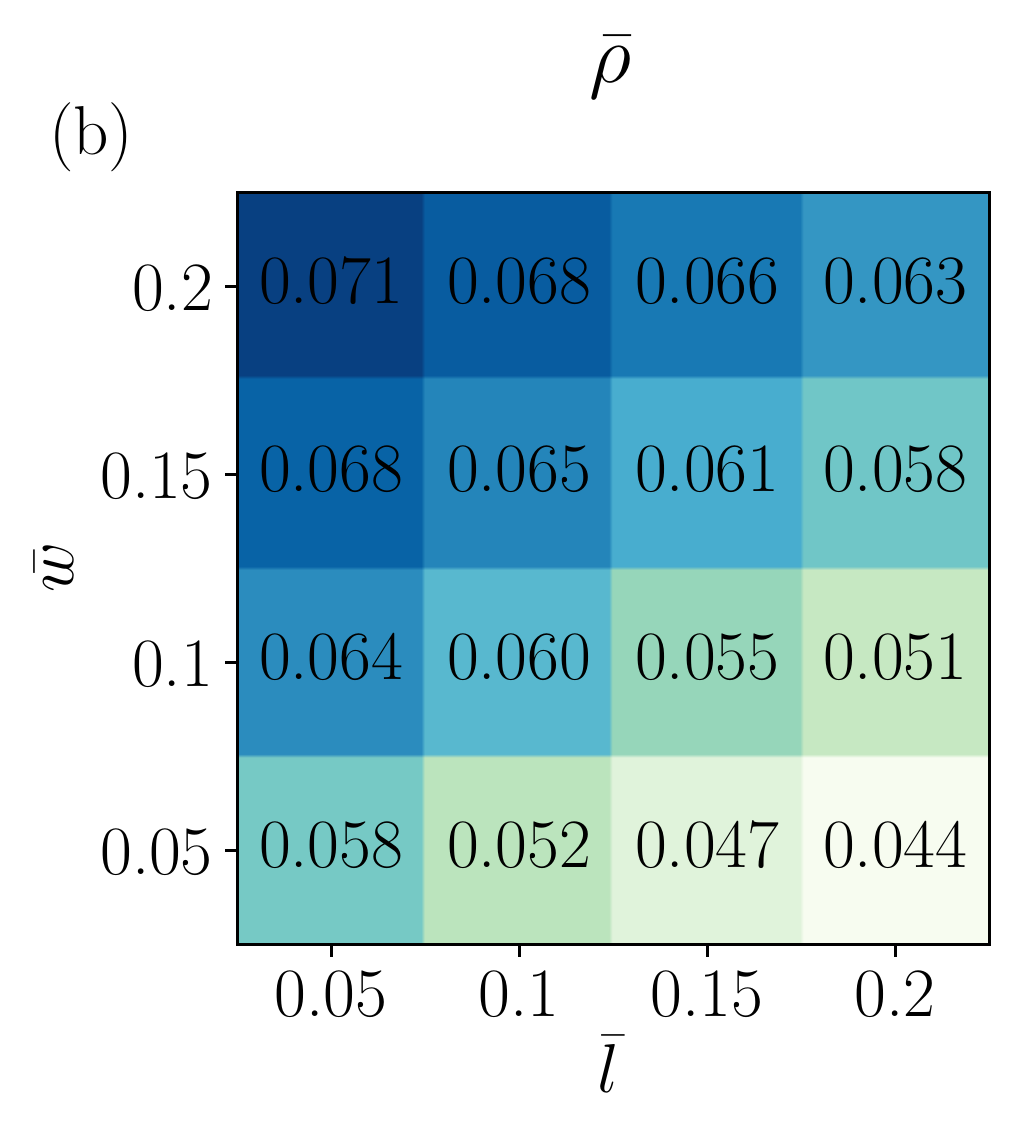}
     \includegraphics[scale=.45]{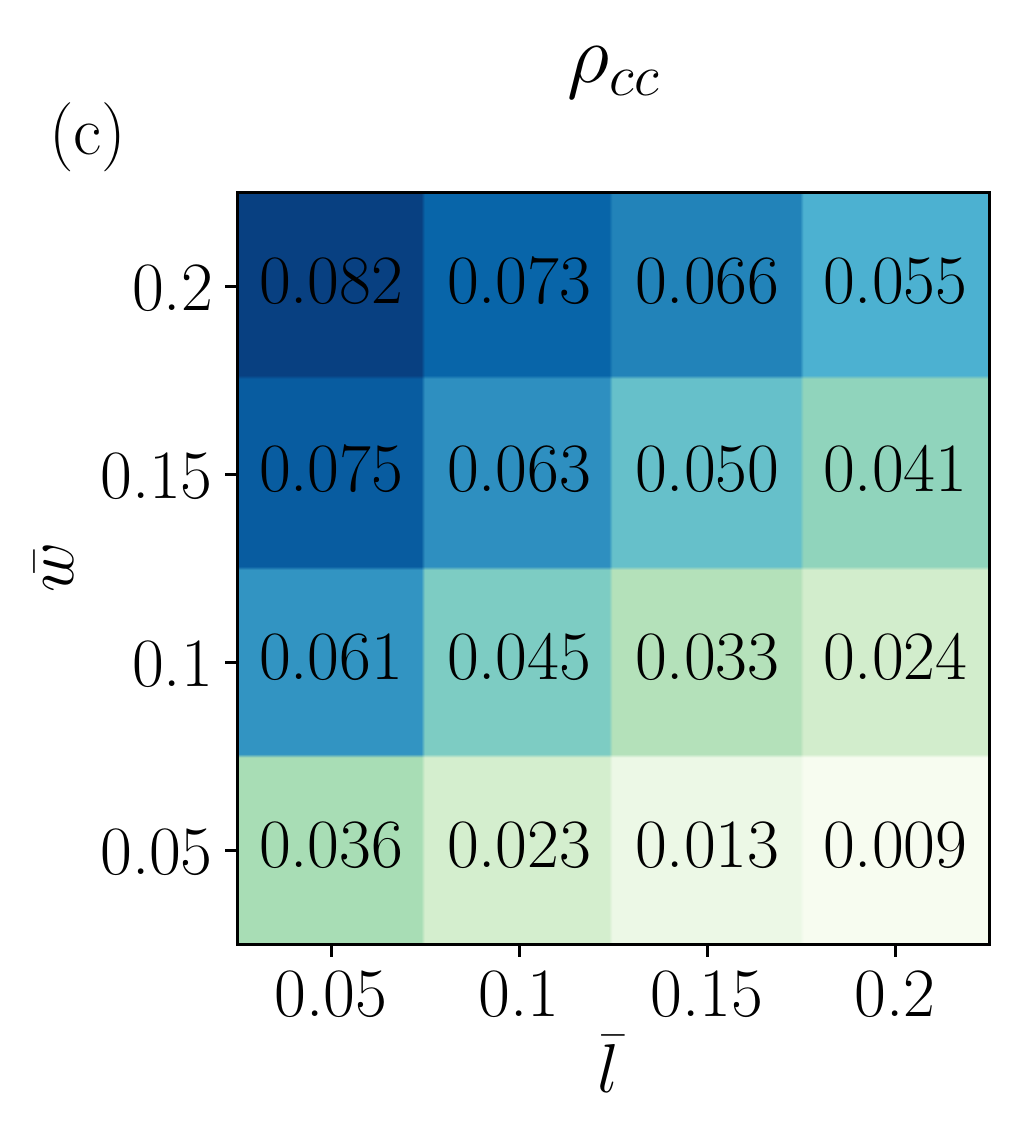}
    \caption{Basic CP: $L_1=L_2=40$ and $\lambda-\lambda_c=0.0512$ ($\lambda=1.70$).(a)  Heat map of the enhancement ratio $e_{\bar w,\bar l}=\tau_{\bar w,\bar l}/\tau_0$;
    (b) Heat map of the average global quasi-stationary density $\bar \rho$, varying $\bar w$ and $\bar l$; (c) Heat map of the density at the center of the corridor $\rho_{cc}$.}
    \label{fig:HMR40}
\end{figure}

We also studied corridors with dimensions: $\bar w=\bar l=0.15$ and $\bar w=0.2$, $\bar l=0.1$. Figure \ref{fig:rCPdL} summarizes the results for the enhancement ratios $e_{\bar w,\bar l}$ for corridor geometries with $0.05\leq \bar w\leq 0.2$ and $0.05\leq \bar l \leq 0.15$ and $L_1=L_2=20$ and $40$. We focus on these corridor sizes  because in real-world applications the goal is often cost-effective restoration of key corridors ensuring conservation without purchasing large areas \cite{Pimm}. As expected from Eq.~(\ref{Eq:tau}),  $\tau$ increases with system size and with $\Delta=\lambda-\lambda_c$ as does the enhancement ratio. Thus, for $L=L_1=L_2=40$ we need a smaller $\Delta$ to obtain the same enhancement as in $L=20$ (see Fig. \ref{fig:rCPdL}). For fixed corridor length $\bar l$ the enhancement increases with  $\bar w$. By contrast, when we increase the length of the corridor for a fixed width $\bar w$ the survival time approaches that of the disconnected regions, as expected (see Fig. \ref{fig:HMR40} (a)).  The global quasistationary density $\bar \rho$ exhibits similar tendencies (see Fig. \ref{fig:HMR40} (b)).

\begin{figure}
    \centering
    \includegraphics[scale=.15]{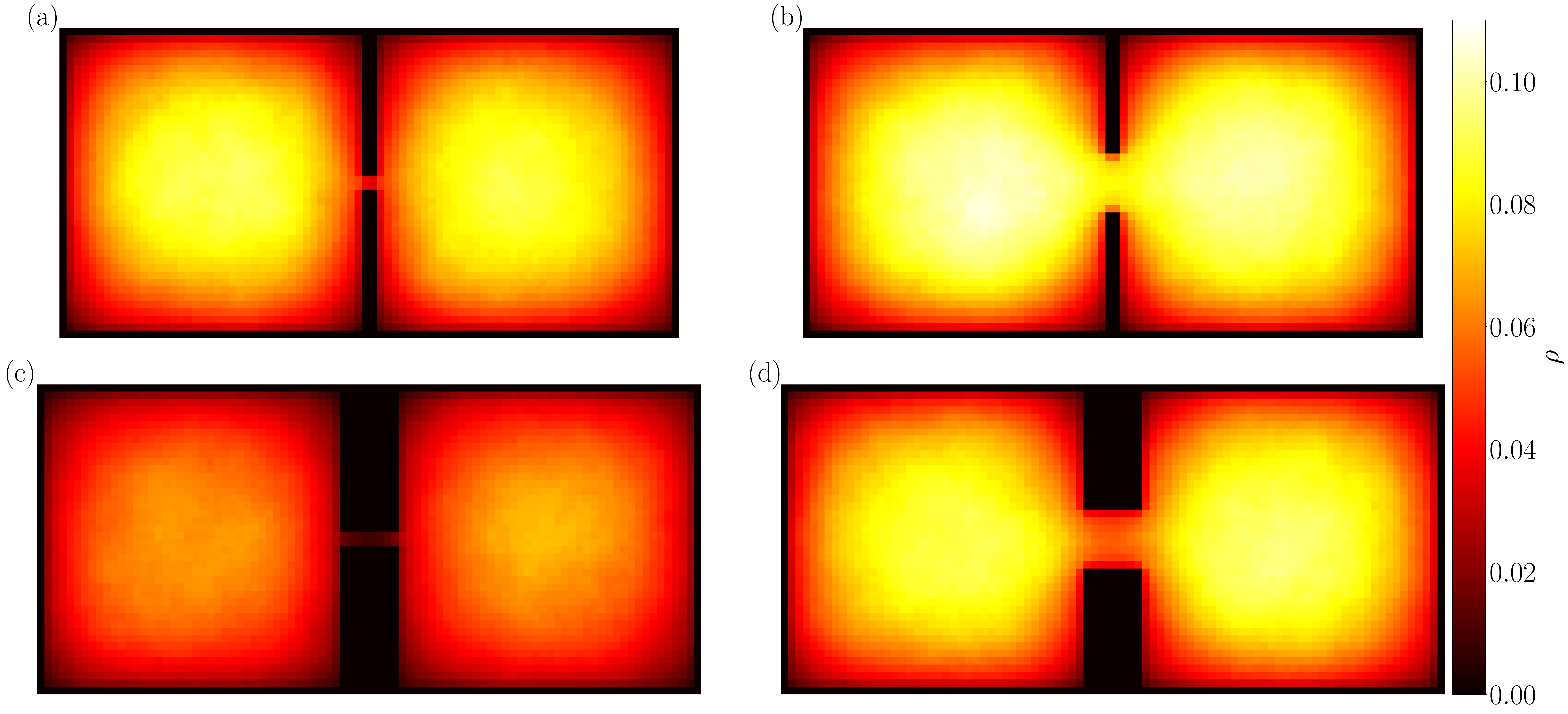}
   
     \caption{Basic CP: Local quasistationary density for $L_1=L_2=40$  and $\lambda-\lambda_c=0.0512$($\lambda=1.70$). (a) $\bar l=0.05$,$\bar w=0.05$; (b) $\bar l=0.05$,$\bar w=0.2$; (c) $\bar l=0.2$,$\bar w=0.05$; (d) $\bar l=0.2$,$\bar w=0.2$.}
     \label{fig:HM}
\end{figure}

Figure \ref{fig:HM} shows heat maps of the local density in the quasistationary state for $L_1=L_2=40$  and $\lambda-\lambda_c=0.0512$ ($\lambda=1.70$), and different corridor geometries: (a) $\bar l=0.05$, $\bar w=0.05$, (b) $\bar l=0.05$, $\bar w=0.2$, (c) $\bar l=0.2$, $\bar w=0.05$ and (d) $\bar l=0.2$,$\bar w=0.2$. The density in the center of the corridor increases with $\bar w$ and decreases with $\bar l$ (see Fig. \ref{fig:HMR40} (c)). The narrower the corridor the lower the density at its center . As expected, the local density is smaller near the borders. For $\bar l=0.2$, $\bar w=0.05$, the density in the corridor is close to zero (0,0087 in its center), and the associated enhancement in lifetime is also negligible. Thus a relatively long narrow corridor represents a barrier for propagation between regions.

\subsection{Diffusive contact process}

\begin{figure}
    \centering
    \includegraphics[scale=.55]{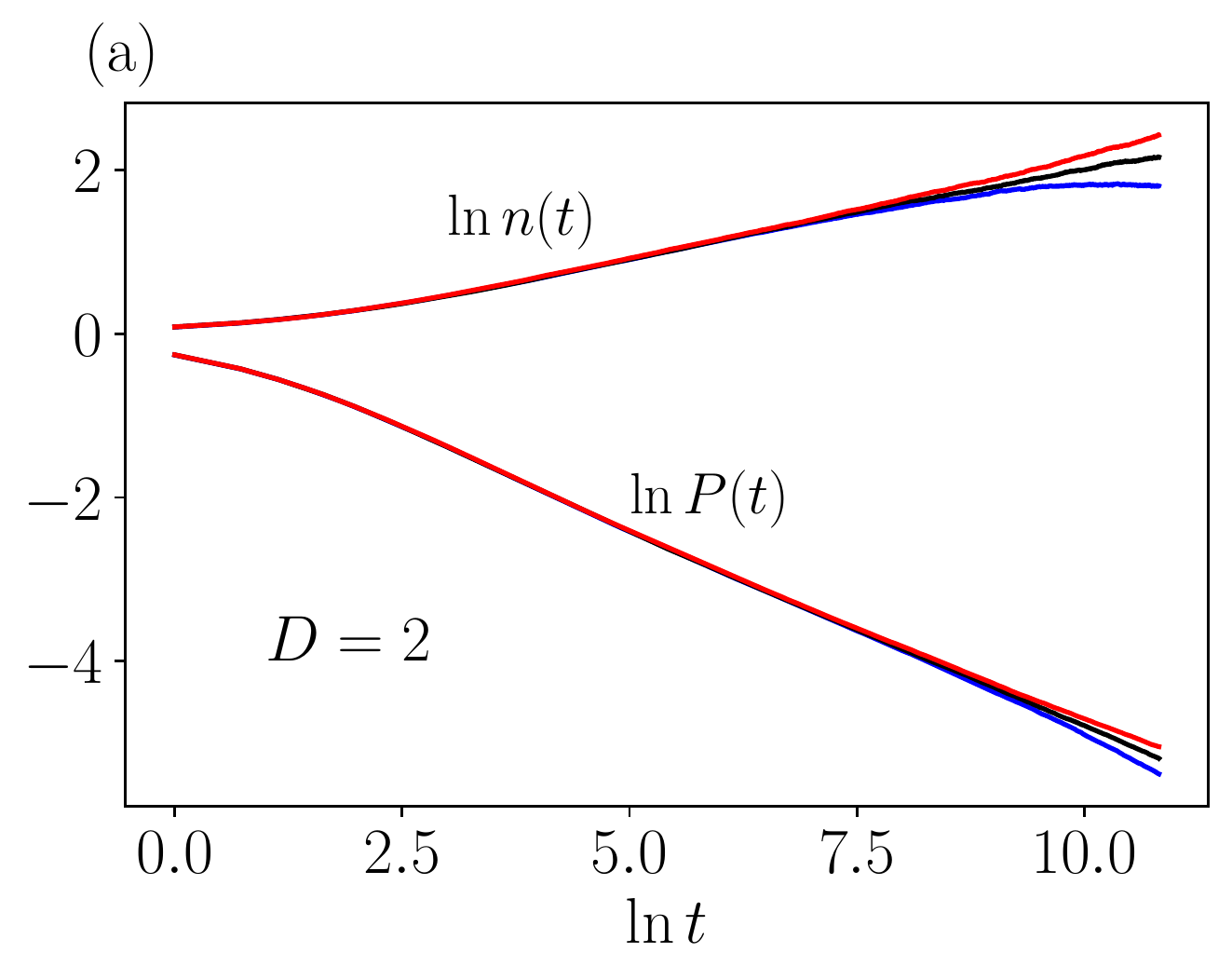}
    \includegraphics[scale=.55]{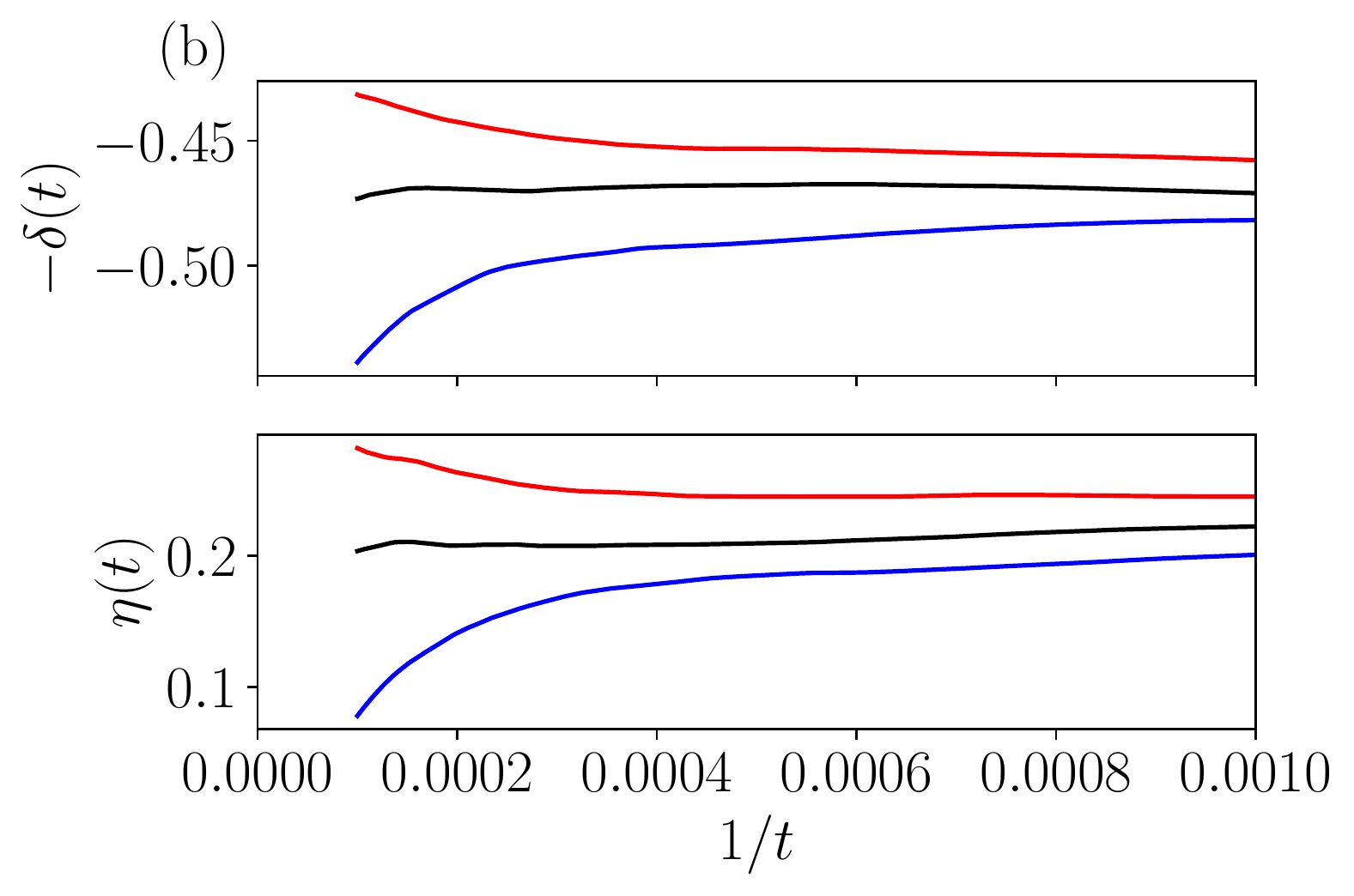}
     \caption{Diffusive CP on the square lattice: (a) survival probability $P(t)$ and number of active sites $n(t)$ in spreading simulations; (b) Local slopes $\delta(t)$ and $\eta(t)$ vs $1/t$. $D=2$ and (from bottom to top) $\lambda=1.3721,1.3723$ and $1.3725$. 
     }
     \label{fig:lambdacD}
\end{figure}

\begin{figure}
    \centering
    \includegraphics[scale=.55]{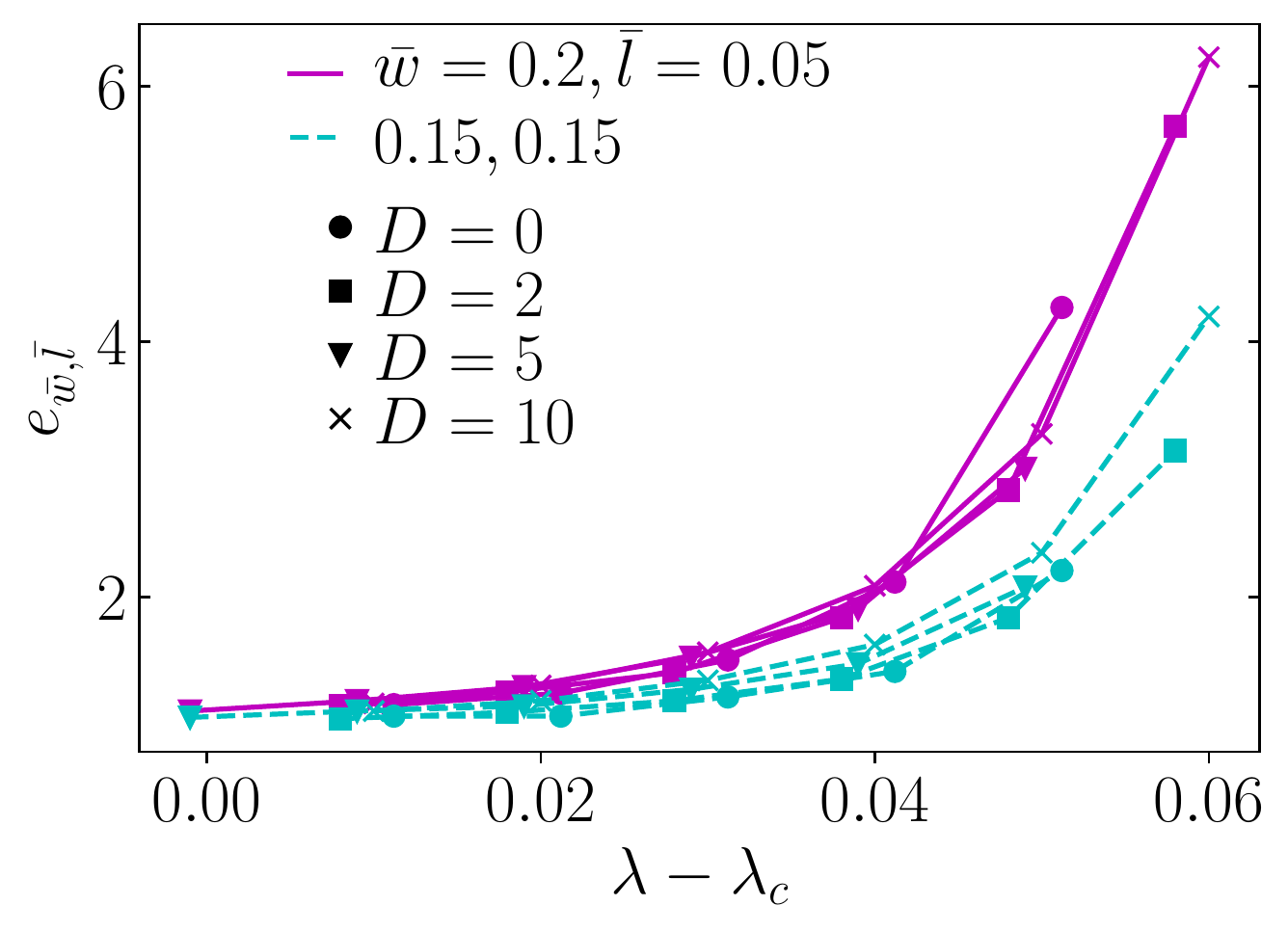}
     \caption{ Diffusive CP: Enhancement ratio $e_{\bar w,\bar l}=\tau_{\bar w,\bar l}/\tau_0$ for different diffusion rates $D$ and two corridor geometries, $\bar w,\bar l$, as indicated. $L_1=L_2=40$. $D=0$ corresponds to the basic CP.}
     \label{fig:DiffCP}
\end{figure}

Since the critical reproduction rate, $\lambda_c(D)$, has not been determined previously for this model on the square lattice, we begin by estimating this value in spreading studies. 
In these studies, performed on a square lattice of size $L=4000$, each realization begins with a single active site, and runs to a maximum time $t_m=5\times10^4$, if the absorbing state is not reached prior to this time; averages are calculated over $5\times10^5$ realizations. At the critical point, one expects power-law behavior of the survival probability $P(t)\sim t^{-\delta}$ and the number of active sites $n(t)\sim t^{\eta}$, where $\delta$ and $\eta$ are critical exponents \cite{Grassberger79, Marro}. Deviations from power law behavior indicate off-critical values of the reproduction rate. Fig. \ref{fig:lambdacD} (a) illustrates this behavior for $D=2$. We obtained precise estimates of $\lambda_c$, $\delta$ and $\eta$ by analyzing the local slopes $\delta(t)$ and $\eta(t)$. $\delta(t)$ and $\eta(t)$ are calculated using linear fits to the data, on the logarithmic scale, on the interval $[t/a,ta]$ (Here $a=5$). Off-critical values are characterized by a curvature in plots of local slopes vs $1/t$. Fig. \ref{fig:lambdacD} (b) shows the behavior of $\delta(t)$ and $\eta(t)$ for $D=2$. The estimates for the spreading exponents are summarized in Table \ref{tab:DiffCP} and are in good agreement with the values for the directed percolation (DP) universality class. 
\begin{table}
    \centering
  \begin{tabular}{c c c c c c}
  \hline \hline
       & $\lambda_c$ &  $\delta$  & $\eta$ \\\hline
    $D=2$ &  1.37235(5) & 0.46670(2) &  0.22343(3)\\
    $D=5$ & 1.24105(5) & 0.45932(3) & 0.20987(7)\\
    $D=10$ & 1.1596(1) & 0.49202(2) & 0.22411(3) \\
    CP or DP & 1.64877(3) & 0.4523(10) & 0.2293(4)\\ \hline \hline
  \end{tabular}   
  
    \caption{Two-dimensional Diffusive CP: results from spreading simulations.
    }
    \label{tab:DiffCP}
\end{table}
With the values of $\lambda_c(D)$ in hand, we can investigate the effect of a corridor on the survival time, following the same procedure as in the basic CP. Figure \ref{fig:DiffCP} shows the enhancement ratio $e_{\bar w,\bar l}$ vs $\lambda-\lambda_c$ for two corridor geometries and four values of the diffusion rate ($D=0,2,5$ and 10) in systems of size $L_1=L_2=40$. $D=0$ corresponds to the basic CP. Increasing the diffusion rate decreases the critical reproduction rate $\lambda_c$, but do not change the enhancement ratio $e_{\bar w,\bar l}$, when plotted as function of $\lambda-\lambda_c$, for the corridor geometries studied here.

\subsection{ Two-species symbiotic contact process}\label{subsec:2SCP}
For the two-species symbiotic contact process (2SCP), we studied two values of the death rate of doubly-occupied sites, $\mu=0.9$ and $0.5$, with critical reproduction rates $\lambda_c=1.64515(5)$ and $\lambda_c=1.47290(5)$, respectively \cite{MMO2012}. $\mu=0.9$ is a relatively small variation of the basic CP($\mu=1$) whereas $\mu=0.5$ represents a rather strong symbiotic interaction. In the 2SCP the behavior of the survival probability $P(t)$ is the same as in the basic CP: after an initial period in which it is equal to one, it decays exponentially with a characteristic survival time $\tau$ (see Fig. \ref{fig:P_CPL1}); the survival time is obtained from linear fits of $\ln P(t)$.

\begin{figure}
    \centering
    \includegraphics[scale=.55]{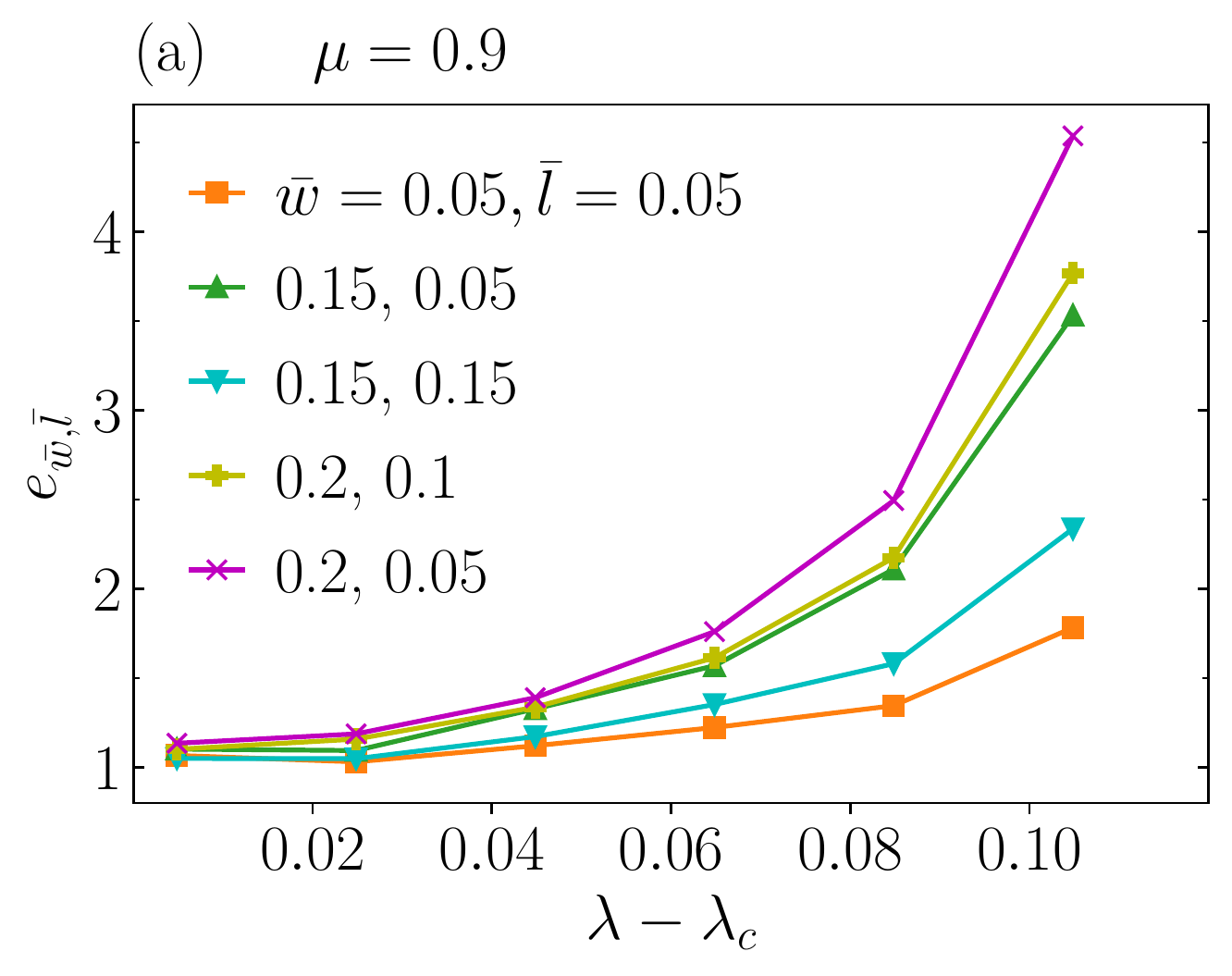}
     \includegraphics[scale=.55]{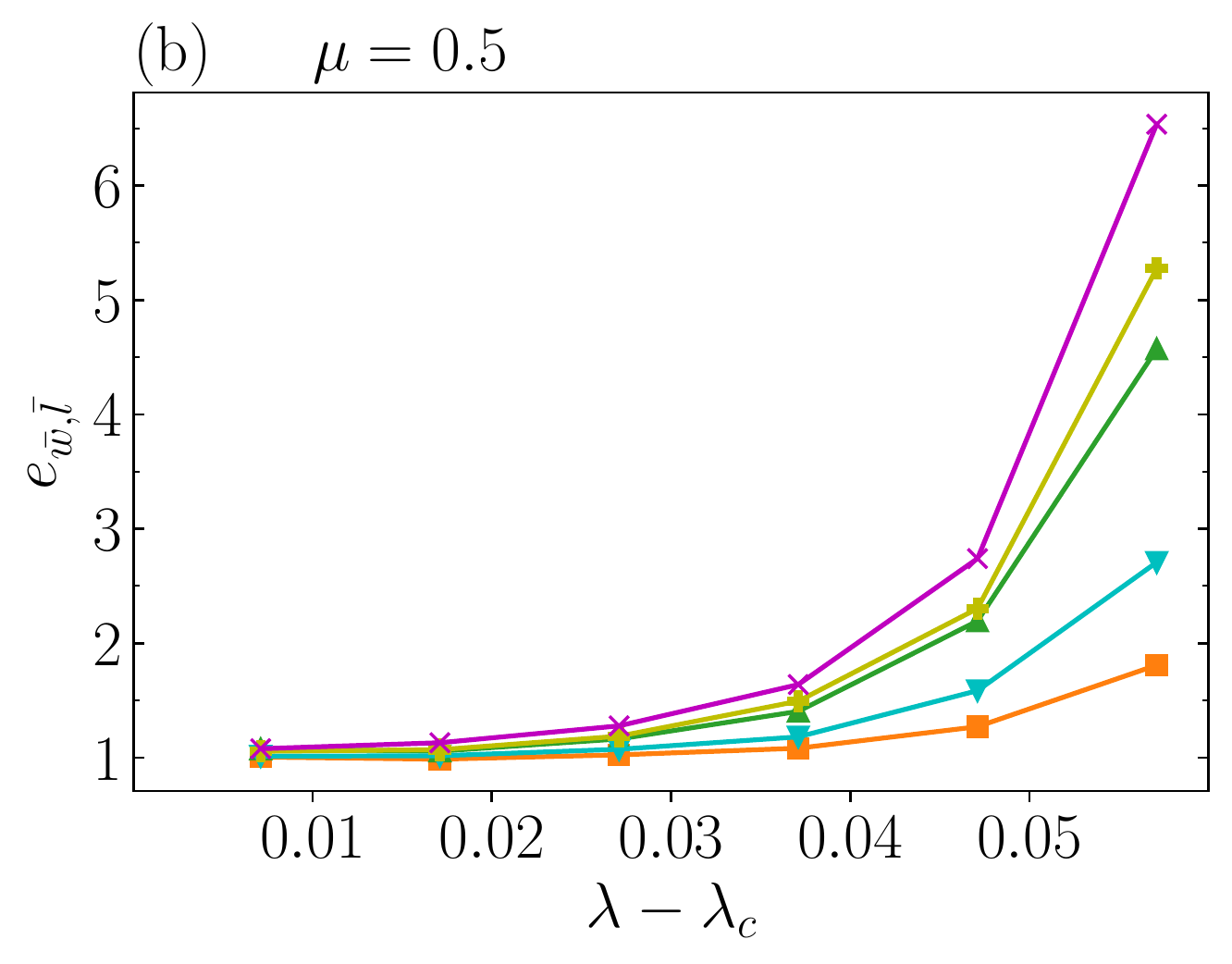}
    \caption{2SCP: Enhancement ratio $e={\bar w,\bar l}=\tau_{\bar w,\bar l}/\tau_0$ for death rates (a) $\mu=0.9$ and (b) $\mu=0.5$. Different symbols indicate different corridor geometries. $L_1=L_2=20$.}
    \label{fig:tau_2SCP}
\end{figure}

Figure \ref{fig:tau_2SCP} shows the enhancement ratio $e_{\bar w,\bar l}$ for $\mu=0.9$ and $\mu=0.5$ and $L_1=L_2=20$. The enhancement in survival time is grater for the stronger symbiotic interaction and exceeds that of the basic CP for the corresponding parameter values (see Table \ref{tab:comp}). 

\begin{table}
    \centering
  \begin{tabular}{c c c c c c}
  \hline \hline
       & CP &  2SCP  & 2SCP \\
       &  & ($\mu=0.9$) & ($\mu=0.5$) \\\hline
    $\tilde{\lambda}$   &  0.0674  & 0.0634 & 0.0388 \\
    $e_{\bar w=0.2,\bar l=0.05}$ & 3.2 & 4.5 & 6.5\\ 
    $e_{\bar w=0.2,\bar l=0.1}$ & 2.8 &  3.8 & 5.3\\ \hline \hline
  \end{tabular}   
  
    \caption{Enhancement ratio $e_{\bar w,\bar l}=\tau_{\bar w,\bar l}/\tau_0$ for the CP and the 2SCP for two regions $L_1=L_2=20$.
    Here $\tilde {\lambda}=(\lambda-\lambda_c)/\lambda_c$).
    }
    \label{tab:comp}
\end{table}

\section{Conclusions}\label{cd}
 Motivated by the use of corridors to counter extinction of species whose habitat is fragmented \cite{Pimm, CE},
we study the effect of connecting regions on the survival time in the basic contact process and two variants, the diffusive contact process, and the two-species symbiotic contact process (2SCP). We compared the survival time of two square regions connected by corridors of width $0<\bar w\le0.2$ and length $0<\bar l\le0.2$ with that of two disconnected regions, i.e., $\bar w=0$. For the three processes short and wide corridors provide the best enhancement of the survival time.

In the diffusive CP, $\lambda_c$ decreases when the diffusion rate $D$ increases. The enhancement in the survival time, when plotted as function of $\lambda-\lambda_c$, does not vary with $D$ for the same corridor geometry and unbiased diffusion. The case in which the diffusion is biased toward the other fragment might be more realistic and will be addressed in future work.

In the 2SCP, increasing the strength of the symbiotic interaction, i.e., decreasing the death rate $\mu$ at sites occupied for both species, decreases $\lambda_c$ and increases the enhancement in the survival time. This suggests that populations with symbiotic interactions enjoy greater benefit when disconnected fragments are linked via corridors.

The contact processes studied here are extreme simplifications of ecological systems but offer a qualitative hint of the impact of corridors on population survival times.  In particular, the spatial structure captures the negative effects of fragmentation such as reduced area and isolation\cite{Haddad15, Ibanez2014}. For the three variants studied, the survival time of the connected regions, $\tau_{\bar w, \bar l}$, is in general greater than that of the disconnected regions, $\tau_0$. For the values of $\lambda$, $L_1$, $L_2$, $\bar w$ and $\bar l$  considered, the  enhancement ratio $e_{\bar w, \bar l}$ ranges from 3 to 7 in the most favorable cases. This result agrees with field observations showing that corridors have a positive impact on populations, aiding in their conservation. Future investigations should address the role of competition between species, e.g., predator-prey dynamics \cite{Tome94} and multispecies models \cite{Pigolotti2018,RMG2021}, as well as the influence of patch arrangement on species richness and persistence \cite{TJORVE2010,Rosch2015,Fahring2017,DELIMAFILHO2021}.

\section{Acknowledgments}
We thank T. V. Rosembach and D. L. P. Lima for sharing their notes and codes, S. Pimm and C.N. Jenkins for helpful communication, and Ricardo Martinez-Garcia and M. M. de Oliveira for helpful comments and discussions. Authors acknowledge financial support from CAPES (I.I. and A.P.F.).
R.D. is grateful to CNPq for financial support under project number 303766/2016-6.

\bibliographystyle{elsarticle-num} 
\bibliography{References}
\end{document}